\documentclass{article}
\usepackage{spconf,amsmath,graphicx,hyperref,url,times,tabularx}
\usepackage[table,xcdraw]{xcolor}
\usepackage{enumitem}
\usepackage{algorithm}
\usepackage{algorithmic}
\usepackage{tablefootnote}
\usepackage{colortbl}

\title{Multi-Domain Audio Question Answering Benchmark Toward \\ Acoustic Content Reasoning}
\name{
      \begin{tabular}{@{}c@{}c@{}}
Chao-Han Huck Yang$^{1}$\quad Sreyan Ghosh$^{1,2}$\quad Qing Wang$^{3}$\quad Jaeyeon Kim$^{4}$\quad Hengyi Hong$^{3}$\quad Sonal Kumar$^{2}$ \\ Guirui Zhong$^{3}$\quad Zhifeng Kong$^{1}$\quad S Sakshi$^{2}$\quad Vaibhavi Lokegaonkar$^{2}$\quad Oriol Nieto$^{5}$ \\Ramani Duraiswami$^{2}$\quad Dinesh Manocha$^{2}$\quad Gunhee Kim$^{4}$\quad Jun Du$^{3}$\quad Rafael Valle$^{1}$\quad Bryan Catanzaro$^{1}$
\end{tabular}}

\address{$^1$NVIDIA \quad $^2$University of Maryland, College Park \\ $^3$University of Science and Technology of China \quad $^4$Seoul National University\quad $^5$Adobe\\
\small \texttt{ hucky@nvidia.com, sreyang@umd.edu, qingwang2@ustc.edu.cn, jaeyeonkim99@snu.ac.kr}
}

\begin{document}
%
\maketitle
\begin{abstract}
We introduce a multi-domain Audio Question Answering (MD-Audio) benchmark designed to advance research in multi-domain sound understanding. The task is organized into three subsets (Bioacoustics, Temporal Soundscapes, and Complex QA), each targeting distinct aspects of interactive question answering over diverse acoustic environments. The benchmark encompasses a broad range of auditory material, including marine mammal vocalizations, environmental soundscapes, and complex real-world audio recordings. We further detail the evaluation protocol, which employs top-1 accuracy alongside an answer-shuffling robustness criterion, and provide baseline results using state-of-the-art audio-language models (Qwen2-Audio-7B, AudioFlamingo 2, and Gemini-2-Flash). The non-archrival benchmark is released as an open-source resource and used for the DCASE 2025 AQA Challenge. Preliminary analyses on the development set reveal substantial variability in model performance across both tasks and domains. Ultimately, this benchmark\footnote{\url{huggingface.co/datasets/PeacefulData/2025_DCASE_AudioQA_Official}} is intended to foster progress in the auditory comprehension and reasoning abilities of audio-language models, capabilities that are critical for enabling systems to perceive, interpret, and interact with the acoustic world in a manner approaching human-level acuity.
\end{abstract}

\section{Introduction}
\label{sec:intro}

Contemporary audio AI research increasingly aims to build systems capable of interactive audio understanding: models that not only recognize sound events but also reason about them \cite{audio_survey}. The MD-Audio is motivated by this goal. It requires systems to interpret diverse acoustic scenes, incorporate relevant external knowledge such as facts or subjective acoustic information about sounds of marine mammals or everyday situations, and reason over both perceived (multiple) sound events and background information to generate appropriate answers.

\begin{figure}[t!]
  \centering
  \includegraphics[width=0.95\linewidth]{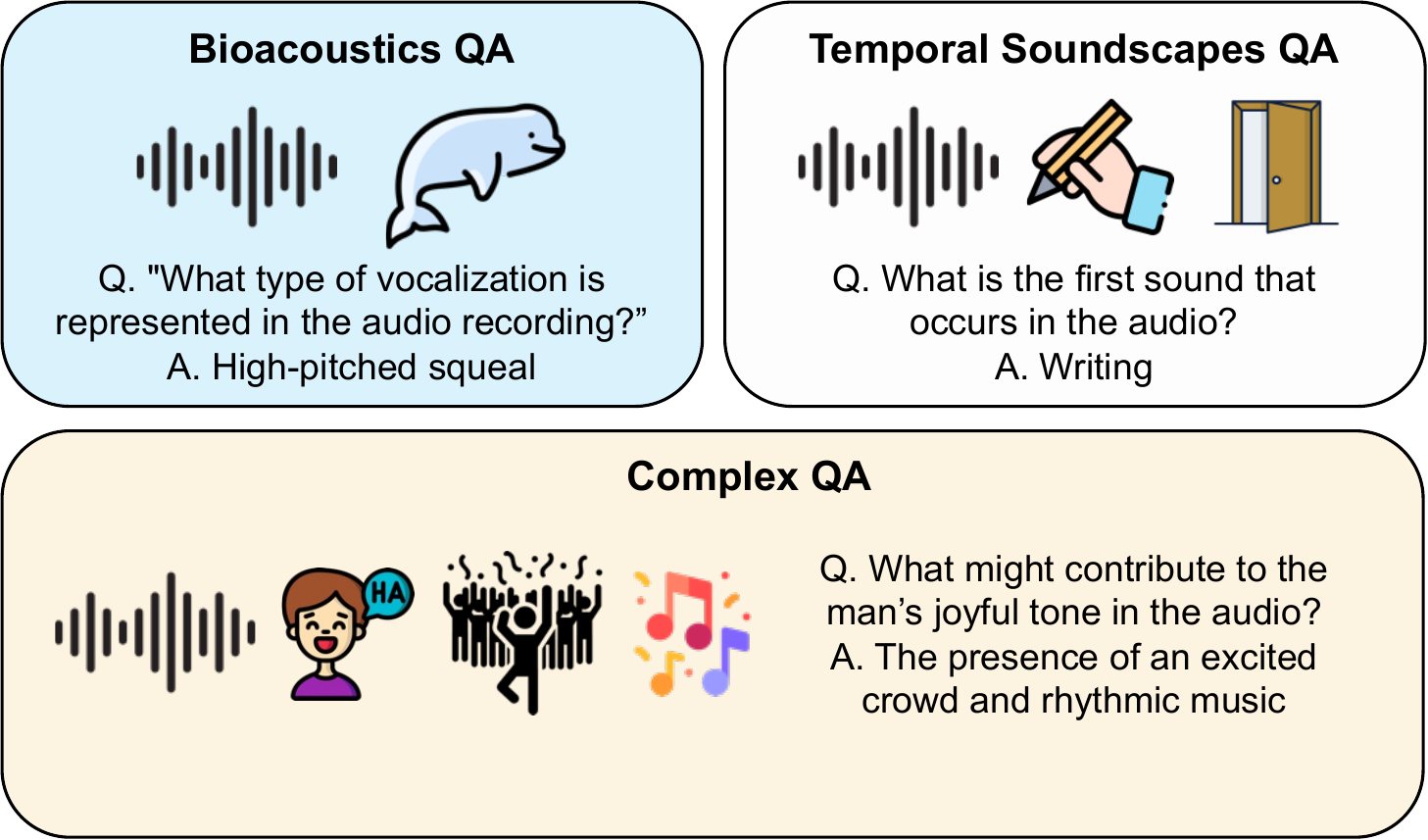}
  \caption{The proposed multi-domain \textbf{audio question answering} (MD-Audio) dataset with three categories: Bioacoustics QA, Temporal Soundscapes QA, and Complex QA.}
  \label{fig:system}
\end{figure}

As illustrated in Figure~\ref{fig:system}, AQA involves scenarios where a listener must both comprehend the audio and make inferences grounded in acoustic and contextual cues.
For example, when asked ``\textit{What might contribute to the man’s joyful tone in the audio?}'', the system must go beyond recognizing the man’s voice. It must also detect background elements, such as the sound of an excited crowd and rhythmic music, and infer that these contextual cues are likely contributing to his joyful tone.
The AQA task spans both general sound events and knowledge-intensive audio content, encouraging comprehensive auditory understanding and multimodal reasoning across a wide range of domains.

Recent advances in large audio-language models \cite{pengi, qwen2audio, salmonn, af1, gama, af2} and the increasing development of benchmarks to assess them \cite{airbench, audiobench, mmau} have underscored the need for evaluation frameworks that better reflect the complexity of audio understanding and reasoning. In line with this trend, AQA is designed as a multi-domain benchmark to probe different facets of audio reasoning. By covering a range of audio domains and question types, the task aims to support progress toward more general and human-like audio intelligence~\cite{vallefugatto}.

\begin{figure}[t]
  \centering
  \includegraphics[width=0.45\textwidth]{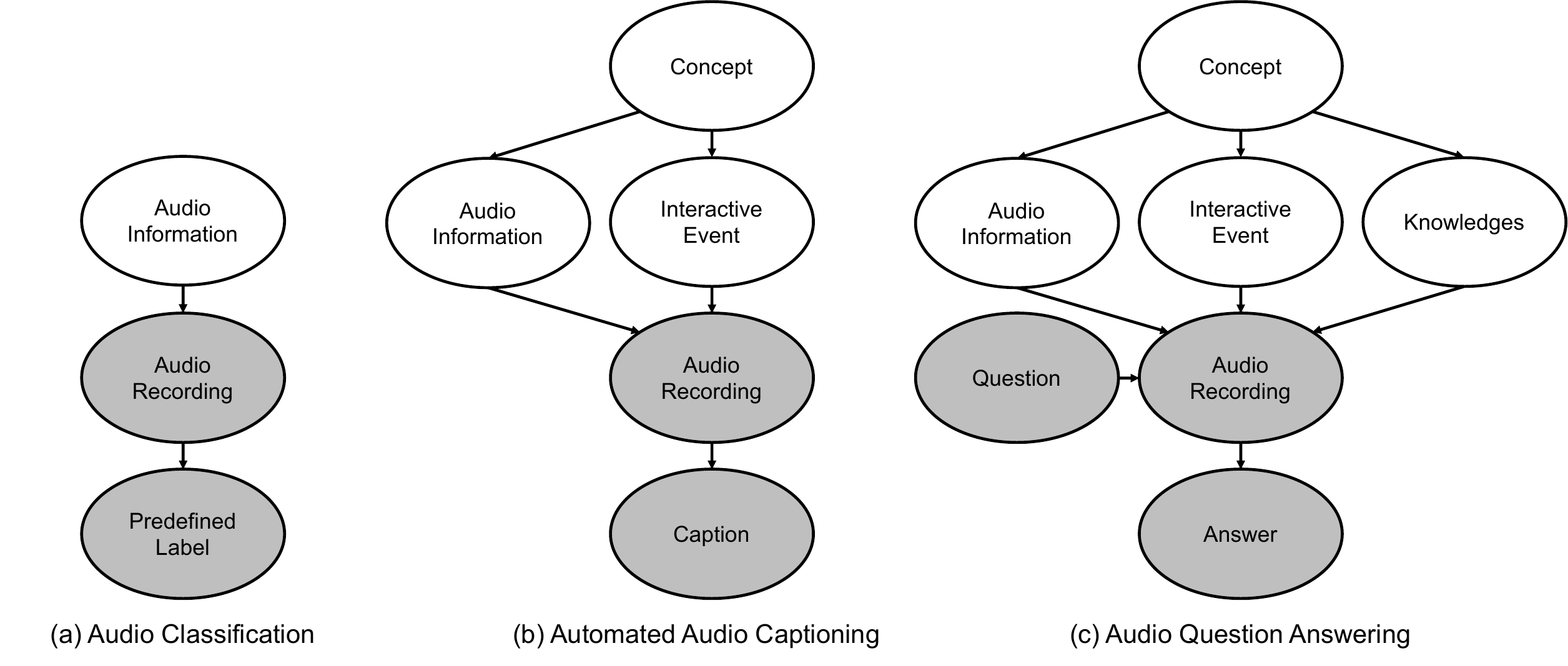}
  \caption{Causal graphs illustrating key differences among three audio understanding tasks. Shaded (gray) nodes represent observable variables, while unshaded nodes represent latent or implicit factors. (a) In Audio Classification, models map audio recordings to predefined labels based solely on surface acoustic features. (b) Automated Audio Captioning introduces an intermediate representation involving interactive events and conceptual understanding to generate natural language descriptions. (c) In Audio Question Answering (AQA), \textbf{reasoning} depends not only on the audio and the question but on latent confounding factors such as audio information, interactive events, and external knowledge. These factors enrich the reasoning process to derive accurate answers.}
  \label{fig:tasks}
\end{figure}

Compared to audio classification tasks~\cite{mesaros2017detection,mesaros2025decade}, which involve understanding audio content and assigning it to a predefined label, AQA requires additional capabilities, including the ability to interpret the question, incorporate relevant knowledge, and link it to the audio information to select the correct answer. A comparison between these tasks is illustrated in Figure~\ref{fig:tasks}. In this study, we present the details of the newly proposed MD-Audio and describe the evaluation methodology, and present baseline model results for the task.


\section{Dataset Details}
The AQA task dataset~\footnote{\url{https://huggingface.co/PeacefulData/2025_DCASE_AudioQA_Baselines}} comprises three distinct QA subsets, each following a multiple-choice question format. Most of the questions present four answer choices (while there are some two and three answer choices in Part 3), with exactly one correct option. 
Each subset is designed to assess a different aspect of audio understanding and reasoning. All subsets include a training set and a development (validation) set for model training and evaluation.

\subsection{Part 1: Bioacoustics QA (BQA)}
 Marine mammals produce a diverse array of acoustic signals for communication, navigation, and other biological functions. These vocalizations tend to be species-specific, enabling fine-grained grounding of audio to real-world biological events. To assess the perceptual and cognitive capabilities of audio-language models, the Bioacoustics QA subset features questions about 31 marine mammal species that vary significantly in vocalization frequency ranges, durations, and ecological habitats. It challenges models to (1) identify the species, vocalization types, or both from audio, and (2) reason about perceived event, especially recalling factual information about the species, interpreting acoustic features, and comparing properties across different sounds.

The subset contains 0.7K training and 0.2K development question-answer pairs. All audio clips are sourced from the Watkins Marine Mammal Sound Database, curated by the Woods Hole Oceanographic Institution and the New Bedford Whaling Museum \cite{watkins}. The recordings span a wide range of conditions, with sample rates from 600 Hz to 160 kHz and durations ranging from 0.4 seconds to over 10 minutes, thereby additionally testing model robustness under diverse acoustic scenarios.

\subsection{Part 2: Temporal Soundscapes QA (TSQA)}

Real-world audio often consists of overlapping or sequential sound events, and understanding their categories, temporal order, and timestamps is critical for temporal reasoning. The Temporal Soundscapes QA dataset is designed to evaluate audio-language models’ ability to perceive and reason about such temporal interactions. It features questions derived from 26 sound classes commonly found in environmental recordings.

This subset challenges models to identify the active sound, classify its type, and infer temporal relationships between events, including their order, onsets, offsets, and durations. The questions are structured with increasing difficulty: for example, identifying the first sound in a clip is relatively straightforward, whereas determining a sound’s duration requires accurately detecting both its onset and offset.

The dataset comprises 1K training and 0.6K development QA pairs. All audio clips are 10 seconds in duration, mono-channel, and sampled at 32–48 kHz. Most audio clips are paired with a single QA example, although a small portion includes up to three related questions. All samples and annotations have undergone careful manual verification.
Audio is sourced from multiple public datasets, including NIGENS \cite{nigens}, L3DAS23 \cite{l3das23}, and TAU Urban Sound 2019 \cite{tau2019}, ensuring a diverse range of acoustic scenes for robust temporal reasoning evaluation.

\subsection{Part 3: Complex QA (CQA)}

Complex QA focuses on complex question answering grounded in audio understanding. Each instance consists of a natural audio clip paired with a multi-faceted question that requires reasoning over temporal, acoustic, and contextual cues within the audio. Questions may involve identifying overlapping sound events, interpreting sequences of auditory phenomena, or discerning abstract relationships implied by the soundscape. This task builds on the principles established in the MMAU Sound subset \cite{mmau}, extending the challenge to higher-order auditory comprehension and inference.

The Complex QA subset comprises 6.4K training and 1.6K development QA pairs, making it the largest among the three subsets. QA pairs are derived from audio clips in AudioSet~\cite{audioset} and the Mira dataset~\cite{mira}. The combined diversity of AudioSet and Mira ensures complex, real-world content suitable for evaluating advanced audio-language reasoning capabilities.



\section{Evaluation}
\subsection{Evaluation Metric}
Submissions will be evaluated on a held-out evaluation set, which will be released on June 1, 2025. The primary evaluation metric is top-1 accuracy, defined as the proportion of questions for which the model’s predicted answer matches the ground truth. Leaderboard rankings will be determined by top-1 accuracy averaged across the three domains (\textit{i.e.}, the mean of the per-domain accuracies) as shown in the row 2 of Tab~\ref{tab:baseline-accuracy}.
If several entries achieve the same score within their \textbf{85\% binomial confidence intervals}, the tie will be broken using their sample-weighted accuracies as shown in the row 3 of Tab~\ref{tab:baseline-accuracy}. From the participants' result, we noticed inference-time strategies (\textit{e.g.}, task-activating prompting \cite{jung2024automatic,yang2023generative}) and policy based post-training adaptation would make model attained competative performance. 


\subsection{Baseline Systems}
\label{sect:4:1}
\noindent\textbf{Qwen2-Audio-7B.} Qwen2-Audio-7B \cite{qwen2audio} is a 7-billion-parameter audio-language model that combines the Whisper-large-v3 audio encoder~\cite{whisper} with the Qwen language model \cite{qwen}. The model was pretrained on a diverse set of audio understanding tasks, including both speech- and non-speech-centric datasets, as well as instruction-following QA involving environmental sounds and multi-turn dialogues. In addition to its base version, the Qwen2-Audio-7B-Instruct variant was further fine-tuned using instruction-tuning datasets designed to improve the model’s ability to follow complex prompts and produce grounded natural language responses conditioned on audio input.

For the AQA task, we use Qwen2-Audio-Instruct to generate answers by providing both the audio clip and the corresponding question prompt. Following the task design, answer choices are included in the input for Parts 1 and 3, as understanding the contextual cues in the options is often essential for arriving at the correct answer. In contrast, for Part 2, where answers are grounded solely in the audio content, the model is queried without access to the answer choices during inference.

As the task requires selecting a single correct option from four candidates, we apply a post-processing step to map the model’s free-form textual output to the most semantically similar option. Specifically, we encode both the model’s response and all candidate answers using a pre-trained Sentence-BERT model \cite{sentence_bert}, compute pairwise cosine similarities, and select the option with the highest similarity score as the final prediction.

\noindent\textbf{AudioFlamingo 2.} AudioFlamingo 2 \cite{af2} is a 3-billion-parameter audio-centric large language model built upon a Flamingo-style cross-attention architecture. It integrates an enhanced CLAP \cite{clap} audio encoder with a language decoder. The model is trained using a multi-stage curriculum that incorporates both real and synthetic data. A key component of this training is the AudioSkills dataset, a large-scale synthetic corpus designed for audio-based instruction following and question answering, covering a few reasoning skills.

For evaluation on the AQA task, we adapted the question format to match the model’s expected input style. Specifically, we reformatted the answer choices from “A. xxx, B. xxx, …” to “(A) xxx. (B) xxx. (C) xxx. (D) xxx.” This modification encouraged the model to generate responses in the standardized form “(A/B/C/D) xxx,” making it easier to interpret outputs.
As the model consistently produced clearly structured answers, we performed answer evaluation via direct string matching against the ground truth, without requiring embedding-based similarity comparisons.

\noindent\textbf{Gemini-2.0-Flash.} Gemini-2.0-Flash \cite{gemini} is a proprietary model developed by Google DeepMind, designed for low-latency, high-efficiency reasoning across multiple modalities. It supports text, image, audio, and video inputs, enabling broad multimodal understanding in real-world scenarios. While its architecture and training details remain undisclosed, Gemini-2.0-Flash accepts long-context audio inputs and demonstrates strong performance on tasks that require audio-conditioned reasoning.

For the AQA task, we prompted Gemini-2.0-Flash with both the audio clip and a unified instruction that includes the question and answer choices. A typical prompt format is: “I will provide you with the question and multiple options. Your task is to generate the only correct option for the question.” For example: “At what time does the first occurrence of the baby crying sound end? The options are: A. 1.9s; B. 3.1s; C. 4.8s; D. 6.0s.”
The model’s output, typically structured as “A. xxx,” is parsed using direct string matching to determine the selected choice, without requiring any embedding-based post-processing.

\begin{table}[t]
\centering
\caption{Zero-shot accuracy (\% with $\pm$ 85\% confidence interval) of baselines on the dev-set across the subsets of the AQA task.}
\label{tab:baseline-accuracy}
\resizebox{\linewidth}{!}{\begin{tabular}{l|ccc}
\hline
\textbf{Dataset} & \textbf{Qwen2-Audio$_{\text{7B}}$} & \textbf{AudioFlamingo2} & \textbf{Gemini-2.0$_{\text{Flash}}$} \\
\hline
domain-avg & 39.6\%($\pm1.9\%$) & 45.0\%($\pm2.0\%$) & \textbf{48.3\%}($\pm2.0\%$) \\
weighted-avg & 45.0\% & 45.7\% & \textbf{52.5\%} \\
\hline
Part 1$_{\texttt{BQA-Dev}}$ & 30.0\% & \textbf{53.9\%} & 42.0\% \\
Part 2$_{\texttt{TSQA-Dev}}$ & 39.2\% & 31.7\% & \textbf{46.3\%} \\
Part 3$_{\texttt{CQA-Dev}}$ & 49.6\% & 49.5\% & \textbf{56.6\%} \\
\hline
\end{tabular}}\label{tab:baseline}
\end{table}

\subsection{Evaluation Results}

The development set accuracy of each model across the three subsets is summarized in Table~\ref{tab:baseline}. Despite being evaluated in a zero-shot setting, overall performance remains low, typically ranging from 30\% to 50\%. This suggests that the proposed AQA dataset probes essential aspects of audio understanding and reasoning that are not effectively addressed by naïve transfer from large-scale pretraining of audio-language models. These results also highlight substantial room for improvement, potentially requiring more sophisticated approaches tailored to the characteristics of each subset.

Moreover, performance varies significantly across models and subsets, underscoring the importance of multi-domain AQA benchmarks in revealing differences in the capabilities of audio-language models. For example, Qwen2-Audio-7B demonstrates moderate performance overall but performs notably poorly on Part~1, which requires fine-grained perception and understanding of acoustic details. In contrast, AudioFlamingo~2 excels in Part~1 but struggles with Part~2, which emphasizes temporal reasoning. Gemini~2~Flash consistently outperforms the other models across all parts. These results highlight the models’ complementary strengths and weaknesses and pose an additional challenge to participants: improving performance across diverse domains simultaneously requires more balanced and generalizable audio understanding and reasoning capabilities.

\subsection{Qualitative Analysis}

This section provides a qualitative evaluation of model performance across three distinct scenarios: hallucinations, correct predictions (wins), and incorrect predictions. Each scenario is presented through carefully selected examples.

\begin{table}[h!]
\centering
\caption{Hallucination responses in Qwen2-Audio-7B }
\begin{tabular}{|p{2cm}|c|c|p{3cm}|}
\hline
\cellcolor[HTML]{EFEFEF}\textbf{Question} & \cellcolor[HTML]{EFEFEF}\footnotesize{\textbf{Correct}} & \cellcolor[HTML]{EFEFEF}\footnotesize{\textbf{Model}} & \cellcolor[HTML]{EFEFEF}\textbf{Model}\textbf{ Explanation (Excerpt)} \\
\hline
Which sounds occur from 2.0s to 4.0s? & A & \cellcolor[HTML]{DAE8FC}B & Mentions ``mechanisms'', ``mechanical fan'', ``respiratory sounds'', ``ticking clock'', none of which match options exactly \\
\hline
What sound overlaps with the keyboard sound? & C & \cellcolor[HTML]{DAE8FC}C & States ``Laughter overlaps'' (not listed among options), although answer matches by chance \\
\hline
\end{tabular}

\label{tab:hallucinations}
\end{table}

\noindent\textbf{Hallucinations.}
Hallucinations in audio-language models refer to instances where the model generates outputs grounded in non-existent acoustic evidence, inventing events or sounds not supported by the input waveform. Table~\ref{tab:hallucinations} presents three representative cases that reveal distinct failure modes in Qwen2-Audio-7B.

In the first example, the model identifies sounds such as ``mechanical fan'' and ``ticking clock'' between 2.0s and 4.0s, none of which are part of the annotated ground truth or aligned with the provided answer choices. This suggests the model is either over-relying on statistical priors or misinterpreting background textures as discrete sound events. This indicates a common issue in models lacking temporal resolution calibration. In the following case, although the model selects the correct answer (``C''), the justification involves an event (``laughter'') that is not among the choices. 


\begin{table}[h!]
\centering
\caption{Correct \& wrong responses in Qwen2-Audio-7B }
\begin{tabular}{|p{2cm}|c|c|p{3cm}|}
\hline
\cellcolor[HTML]{EFEFEF}\textbf{Question} & \cellcolor[HTML]{EFEFEF}\footnotesize{\textbf{Correct}} & \cellcolor[HTML]{EFEFEF}\footnotesize{\textbf{Model}} & \cellcolor[HTML]{EFEFEF}\textbf{Model}\textbf{ Explanation (Excerpt)} \\
\hline
Which acoustic characteristics describe the signal? & B & \cellcolor[HTML]{67FD9A}B & Matches exactly: ``rapidly modulating pattern between 10--20 kHz''\\
\hline
What is the start time of the cupboard sound? & C & \cellcolor[HTML]{FFCCC9}B & States 8.1s, which does not match any correct option (6.5s) \\
\hline
\end{tabular}
\label{tab:wins}
\end{table}

\noindent\textbf{Correct and Wrong Predictions .}
The model accurately identifies fine-grained acoustic patterns and species-specific vocalizations. Notably, it demonstrates precise spectral reasoning (e.g., identifying modulated high-frequency components). Incorrect predictions arise when the model either misattributes acoustic sources or fails to temporally align events with their corresponding semantic categories.



\section{Conclusion}
\label{sec:con}
\vspace{-0.1cm}

We introduce the AQA task for the DCASE 2025 Challenge, which requires systems to answer multiple-choice questions based on audio inputs. The task comprises three subsets, Bioacoustics QA, Temporal Soundscapes QA, and Complex QA, each targeting different aspects of audio understanding and reasoning. We outline top-1 accuracy for ranking and permutation-based accuracy for robustness. 


\bibliographystyle{IEEEbib}
\bibliography{strings,refs}

\end{document}